\newcommand\aap{{A}\&{A}}

\newcommand\aj{{AJ}}
\newcommand\apj{{ApJ}}

\newcommand\apjs{{ ApJS}}

\newcommand\mnras{{MNRAS}}

\newcommand\nature{{Nature}}
\newcommand\pasp{{PASP}}

\documentclass[useAMS,usenatbib]{mn2e}
\usepackage{graphicx}
\usepackage{epsf}
\usepackage{bm}
\usepackage{color}
\usepackage{times}
\title{{\it HST\/} and LAMOST discover a dual active galactic nucleus in J0038+4128}

\author[Y. Huang et al.]
               {Y. Huang$^{1}$\thanks{E-mail: yanghuang@pku.edu.cn},  
                 X.-W. Liu$^{1,2}$\thanks{E-mail: x.liu@pku.edu.cn}, H.-B. Yuan$^{2}$, M.-S. Xiang$^{1}$, Z.-Y. Huo$^{3}$ , Y.-H. Hou$^{4}$, G. Jin$^{5}$, 
                 \newauthor Y. Zhang$^{4}$, X.-L. Zhou$^{3}$\\
$^{1}$Department of Astronomy, Peking University, Beijing 100871, People's Republic of China.\\
$^{2}$Kavli Institute for Astronomy and Astrophysics, Peking University, Beijing 100871, People's Republic of China\\
$^{3}$National Astronomical Observatories, Chinese Academy of Sciences, Beijing 100012, People's Republic of China\\
$^{4}$Nanjing Institute of Astronomical Optics \& Technology, National Astronomical Observatories, Chinese Academy of Sciences, Nanjing 210042, People's Republic of China\\
$^{5}$University of Science and Technology of China, Hefei 230026, People's Republic of China}

\begin{document}

\date{}

\pagerange{\pageref{firstpage}--\pageref{lastpage}} \pubyear{2013}

\maketitle

\begin{abstract}
We report the discovery of a kiloparsec-scale dual active galactic nucleus (AGN) in J0038+4128. 
From the {\it Hubble Space Telescope\/} ({\it HST\/}) Wide Field Planetary Camera 2 (WFPC2) images, we find two optical nuclei with a projection separation of 4.7 kpc (3{\farcs}44).
The southern component  (J0038+4128S) is spectroscopically observed with the {\it HST\/} Goddard High Resolution Spectrograph in the UV range and is found to be a Seyfert 1 galaxy with a broad Ly$\alpha$ emission line.
The northern component (J0038+4128N) is spectroscopically observed during the Large Sky Area Multi-Object Fibre Spectroscopic Telescope (also named the Guoshoujing Telescope) pilot survey in the optical range. 
The observed line ratios as well as the consistency of redshift of the nucleus emission lines and the host galaxy's absorption lines indicate that J0038+4128N is a Seyfert 2 galaxy with narrow lines only.
These results thus confirm that J0038+4128 is a Seyfert 1-Seyfert 2 AGN pair.
The {\it HST\/} WFPC2 $F$336$W$/{\it U\/}-band image of J0038+4128  also reveals for the first time for a dual AGN system two pairs of bi-symmetric arms, as are expected from the numerical simulations of such system.
Being one of a few confirmed kiloparsec-scale dual AGNs exhibiting a clear morphological structure of the host galaxies, J0038+4128 provides an unique opportunity to study the co-evolution of the host galaxies and their central supermassive black holes undergoing a merging process.

\end{abstract}


\section{Introduction}
\label{sec:optaxrat_int}
In the hierarchical $\Lambda$ cold dark matter ($\Lambda {\rm CDM}$) cosmology, galaxies built up via mergers (Toomre \& Toomre 1972).
Binary supermassive black holes (SMBHs) are natural outcomes of galaxy mergers (Begelman et al. 1980; Milosavljevi\'c \& Merritt 2001; Yu 2002), since almost all massive galaxies are believed to host a central SMBH. 
In the gas-rich case, the strong tidal interactions caused by galaxy mergers can trigger the active galactic nucleus (AGN) by sending a large amount of gas to the central regions (Hernquist 1989; Kauffmann \& Haehnelt 2000; Hopkins et al. 2008).
A dual AGN could emerge if the two merging SMBHs are both simultaneously accreting gas in a gas-rich major merger.
Finding dual AGNs, especially those exhibiting two black holes on a kiloparsec-scale (Liu et al. 2013), will provide important clues to understand the relation between the AGN activity and the galaxy evolution and AGN physics (Colpi \& Dotti 2011; Yu et al. 2011).  

Over the past few years, hundreds of dual AGNs of separations greater than 10 kpc have been discovered (e.g. Myers et al. 2007, 2008; Green et al. 2010; Piconcelli et al. 2010). 
However, such systems represent only the earliest stage of the binary SMBH evolution.
No more than dozens of close dual AGNs (separations between 1 and 10 kpc) are found both from either systematic searches (e.g. Liu et al. 2010, 2013; Rosario et al. 2011) or by serendipitous discoveries (e.g. Junkkarinen et al. 2001; Komossa et al. 2003; Ballo et al. 2004; Bianchi et al. 2008; Fu et al. 2011; Koss et al. 2011; McGurk et al. 2011).  
However, most of them only show two prominent nuclei either in the radio, X-ray or the optical band, the properties of the host galaxies are poorly known. The latter are extremely important for probing the relation between the galaxy evolution and nucleus activity (Shields et al. 2012).

Here, we report the discovery of a new dual AGN in J0038+4128 ({\it z\/} = 0.0725) with a spatial separation of 4.7 kpc (3\farcs44). A $F$336$W$/{\it U\/}-band image of J0038+4128 obtained with the Wide Field Planetary Camera 2 (WFPC2) on board the {\it Hubble Space Telescope\/} ({\it HST\/}) reveals for the first time for a dual AGN system two pairs of bi-symmetric arms.
In Section 2.1, the {\it HST\/} photometric and spectroscopic observation and data reduction of J0038+4128 are introduced.
We describe the Large Sky Area Multi-Object Fibre Spectroscopic Telescope (LAMOST; also known as the Guoshoujing Telescope)  spectroscopic observation and data reduction in Section 2.2.
The results are  discussed and concluded in Section 3, followed by a brief summary in Section 4.
We adopt a $\Lambda$CDM cosmology with $\Omega_{\rm m}$=0.3, $\Omega_{\Lambda}$=0.7, and $H_0$= 70 km s$^{-1} $Mpc$^{-1}$ throughout. 
All quoted wavelengths are in vacuum units.

\section{OBSERVATIONS AND REDUCTIONS}

\subsection{{\it HST} Images and Spectra}

The images presented here were obtained with the WFPC2 on board the {\it HST\/}\footnote{Based on observations made with the NASA/ESA {\it Hubble Space Telescope}, and obtained from the Hubble Legacy Archive, which is a collaboration between the Space Telescope Science Institute (STScI/NASA), the Space Telescope European Coordinating Facility (ST-ECF/ESA) and the Canadian Astronomy Data Centre (CADC/NRC/CSA).} as parts of the General Observer (GO) programme GO-6749 (PI: Laura Danly) on 1996 August 27. WFPC2 contains four Loral CCD detectors of 800$\times$800 pixels. 
The field of view (FOV) of the Planetary Camera is about 34$\times$34 arcsec$^{2}$ (0\farcs046 pixel$^{-1}$) , whereas that of each of the three Wide Field (WF) arrays  is 150$\times$150 arcsec$^{2}$ (0\farcs1 pixel$^{-1}$).
Fig.\,1 shows the WFPC2 combined images of J0043+4128 (located in one of the three WFs) of two deep exposures (600\,s) in the $F$336$W$/{\it U\/}-band and two deep exposures (400\,s) and one shallow exposure (30\,s) in the $F$555$W$/{\it V\/}-band.
The combined images were produced at the Canadian Astronomical Data Center (CADC) with MultiDrizzle with improved astrometry and geometric distortions correction\footnote{http://hla.stsci.edu/}.
J0038+4128 is well resolved into two components, a main (S) and a companion (N), separated by 3\farcs44\, in both bands.

The UV spectrum of J0038+4128S was obtained with the {\it HST\/} Goddard High Resolution Spectrograph (GHRS) on 1997 February 4, also under project GO-6749. 
Six exposures were taken with the G140L grating through the Large Science Aperture (2\farcs0$\times$2\farcs0) for a total integration time of 18,822 s.
The resolving power is about 1000 at 1200 {\AA}, which corresponds to $\sim$300 km s$^{-1}$.
The spectra  were processed using the standard CALHRS v1.3.14 reduction pipeline with the latest reference files. 
We compute the wavelength zero-points of the spectra with the nearest SPYBAL\footnote{SPYBAL stands for SPectrum Y BALance, and it is performed to ensure proper alignment of the spectrum on the science diodes.} calibration spectrum  observations in time using the STSDAS IRAF task {\it waveoff} and find offsets ranging from 0.6956 to 0.9948 {\AA} for individual exposures.
After correcting for the wavelength zero-points for each exposure we present the final combined spectrum in Fig.\,2.
The spectrum was rebinned to a common wavelength grid with a constant step size of 0.57 {\AA}. 

\subsection{LAMOST spectra}
\begin{table*}
    \centering
    \caption{Observation}
    \begin{tabular}{ccccccccc}
    \hline
     Area  & RA & Dec. & Facility  & Date & Seeing&  Exp. time&Spectral coverage & Resolving power\\
    (c.f. Fig. 1) & & & & &(arcsec) &(second)&(\AA)&\\
    \hline
     Yellow circle&00 38 33.05 &+41 28 53.48&LAMOST   &2011 Oct 05&4.0&1800 &3700--9000 &1800\\
     Red      circle&00 38 33.36 &+41 28 51.95&LAMOST  &2011 Oct 24&3.4&3600 &3700--9000  &1800\\
     Red      circle&00 38 33.36 &+41 28 51.95&LAMOST  &2011  Oct 28&2.8&2100 &3700--9000  &1800\\
     Yellow     box &00 38 33.12 &+41 28 50.30&{\it HST\/}/GHRS&1997  Feb 04&---&18,822&1275--1561&1000\\
     \hline
     \end{tabular}
\end{table*}

The optical spectrum of J0038+4128N were observed during the LAMOST pilot survey on 2011 October 5.  
LAMOST is a 4 metre quasi-meridian reflecting Schmidt telescope equipped with 4000 fibres, each of an angular diameter of 3\farcs3 projected on the sky,  in an FOV of 5$^{\circ}$ in diameter (Cui et al. 2012). The spectra has a resolution of {\it R} $\sim$ 1800 and covers wavelengths from 3700 to 9000 {\AA}.
In addition to J0038+4128N, LAMOST also observed a region a few arcsec east of the two nuclei marked as J0038+4128H shown in Fig.\,1 on 2011 October 24 and 28. 
The weather of the three nights was clear but of relatively poor seeing ($\sim$3\arcsec).
Long-term monitoring of the LAMOST fibre positioning accuracy  (Yuan et al., in preparation) shows that it varies from fibre to fibre. 
However, the observations of J0038+4128N and J0038+4128H were obtained with fibres of the highest precision (better than 0\farcs5).
The data were processed using the LAMOST standard pipeline, with flux calibration better than $\sim$ 10\% (Liu et al. 2013).

We summarize the details of all spectral observations of J0038+4128 in Table 1, including the three positions of  the observed regions, central positions of aperture, telescopes used, observational dates, seeing, total exposure times, spectral coverage and resolving power.
In addition, the photometric properties of J0038+4128 from archival observations are summarized in Table 2.

\begin{figure}
  \centering
  \includegraphics[width=3.75cm,height=5.5cm]{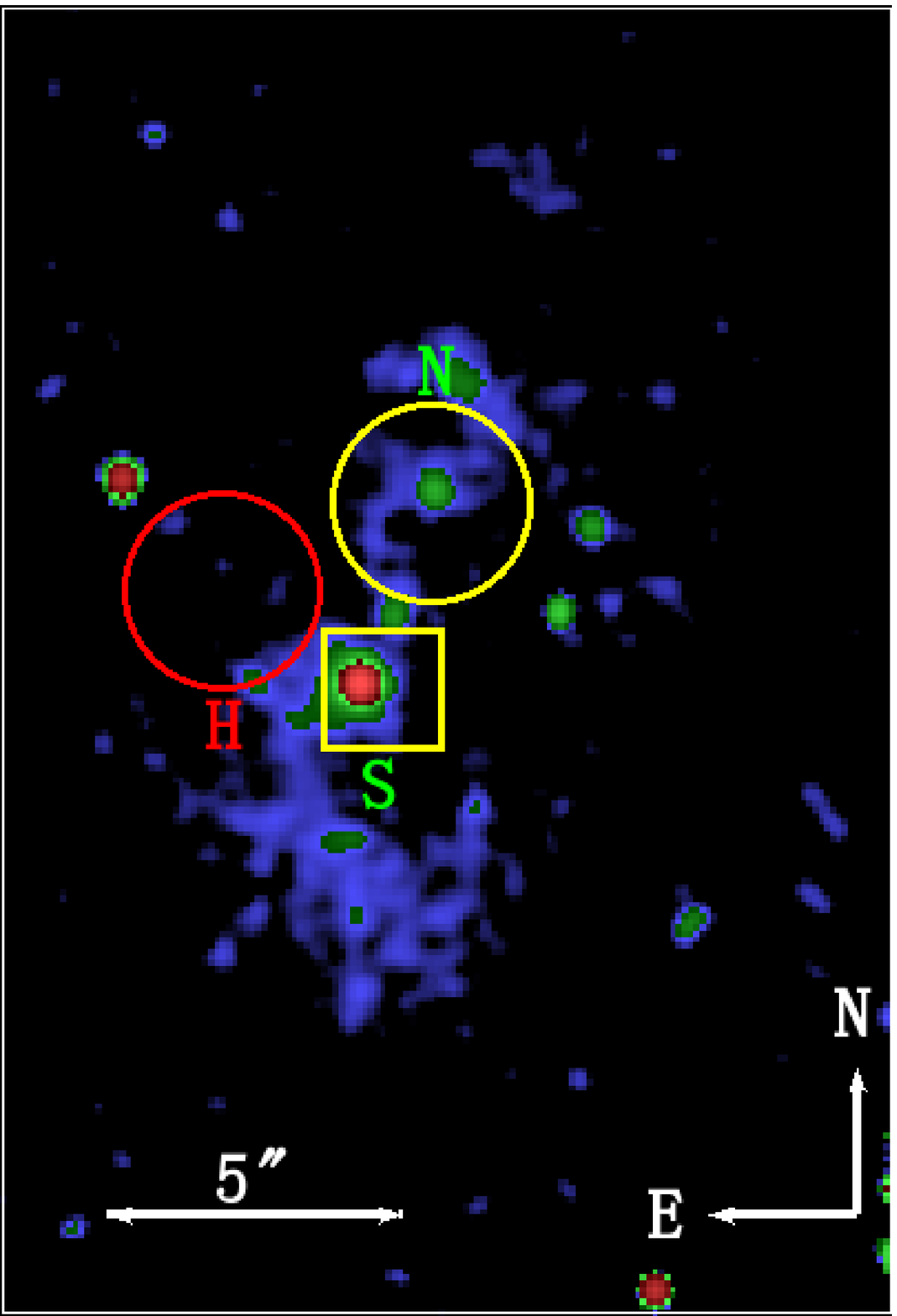}
  \includegraphics[width=3.75cm,height=5.5cm]{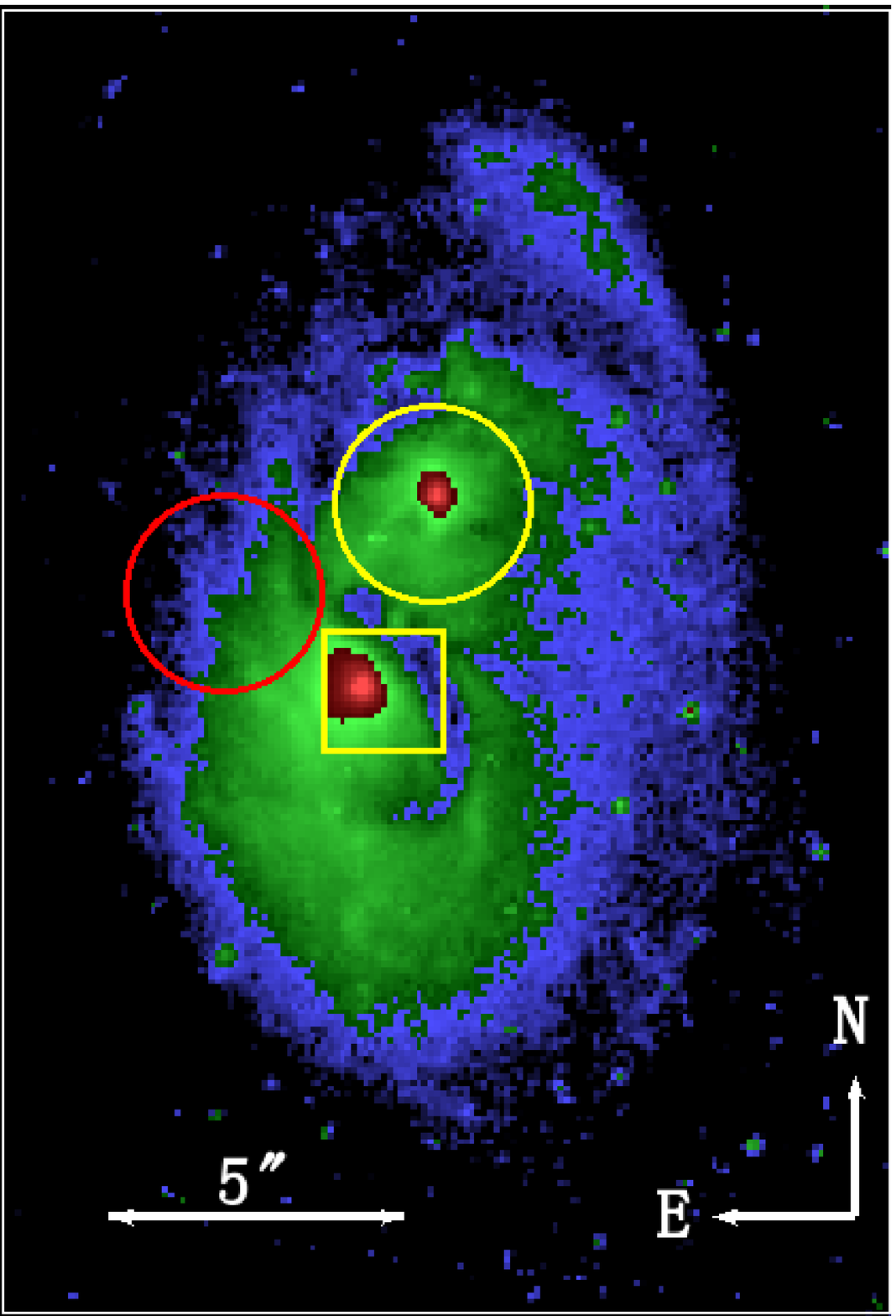}
\caption{Left: {\it HST\/}/WFPC2 $F$336$W$/{\it U\/}-band pseudo colour image of J0038+4128 on a logarithmic scale (smoothed using the Gaussian function available in SAOImage DS9 6.2). Right: {\it HST\/}/WFPC2 $F$555$W$/{\it V\/}-band pseudo colour image of J0038+4128 on a logarithmic scale. The two nuclei are clearly resolved in both bands  into two components, the S and the N components. North is up and east is to the left. Spatial scale is also shown in each panel. Apertures used to obtain the spectra presented in this work are marked, including one 2\arcsec $\times$ 2\arcsec rectangular aperture for the {\it HST\/}/GHRS spectroscopy, three circular apertures of 3\farcs3 diameter for the LAMOST spectroscopy. See  Table 1 for detail.}
\end{figure}

\begin{figure*}
  \centering
  \includegraphics[width=15cm,height=7.5cm]{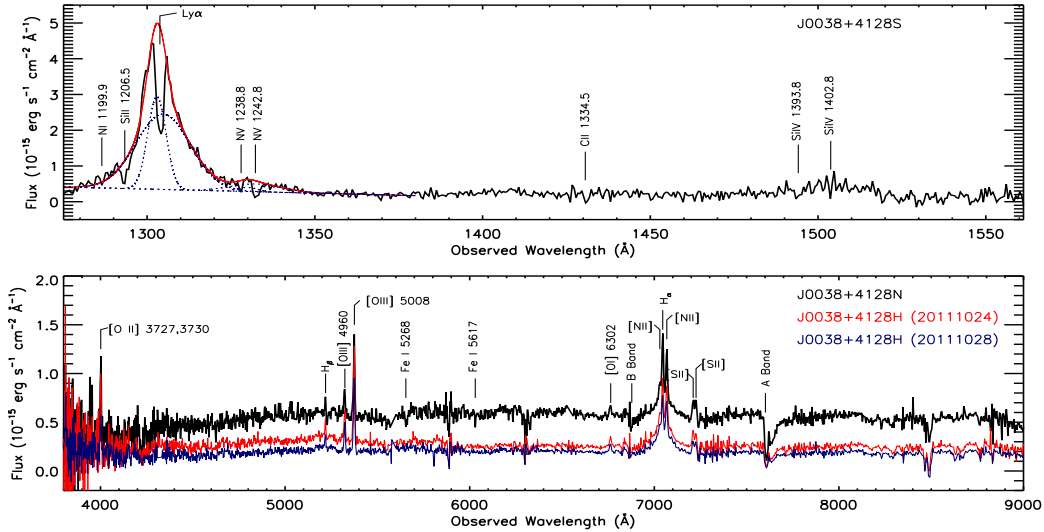}
\vskip-0.25truein
\caption{Top: the UV spectrum of J0038+4128S obtained with the {\it HST}/GHRS  using the G140L grating.   Bottom: LAMOST spectra of J0038+4128N (black) and of J0038+4128H (red and blue). Detected emission lines as well as absorption lines are labelled for J0038+4128S. For J0038+4128N and J0038+4128H, emission lines detected as well as the atmospheric ($A$- and $B$-bands) are marked. In addition, two absorption lines of neutral iron from the host galaxy have been well detected in the spectrum of J0038+4128N. Most other absorption line-like features are artefacts produced by poor sky extraction.}
\end{figure*}

\begin{table}
    \begin{center}
    \caption{Photometric data for J0038+4128}
    \begin{tabular}{ccccc}
    \hline
     Band &$\lambda$ ($\mu$m) & J0038+418S & J0038+418N&Facility\\
     \hline
     FUV&0.16&\multicolumn{2}{c}{19.041$\pm$0.136}&{\it GALEX}\\
     NUV&0.23&\multicolumn{2}{c}{18.469$\pm$0.072}&{\it GALEX}\\
     $u^{a}$&0.36&16.889$\pm$0.015&24.635$\pm$0.975&SDSS\\
     $g^{a}$&0.47&15.604$\pm$0.003&19.985$\pm$0.020&SDSS\\
     $r^{a}$&0.62&14.831$\pm$0.003&18.752$\pm$0.011&SDSS\\
     $i ^{a}$&0.75&14.261$\pm$0.003&18.386$\pm$0.012&SDSS\\
     $z^{a}$ &0.89&13.943$\pm$0.005&18.004$\pm$0.023&SDSS\\
     $J^{b}$&1.25&\multicolumn{2}{c}{13.286$\pm$0.048}&2MASS\\
     $H^{b}$&1.65&\multicolumn{2}{c}{12.556$\pm$0.070}&2MASS\\
     $K^{b}$&2.17&\multicolumn{2}{c}{12.103$\pm$0.074}&2MASS\\
     $W$1&3.40&\multicolumn{2}{c}{11.624$\pm$0.023}&{\it WISE}\\
     $W$2&4.60&\multicolumn{2}{c}{11.070$\pm$0.020}&{\it WISE}\\
     $W$3&12.00&\multicolumn{2}{c}{8.291$\pm$0.019}&{\it WISE}\\
     $W$4&22.00&\multicolumn{2}{c}{5.636$\pm$0.033}&{\it WISE}\\
     \hline
     \end{tabular}
     \end{center}
     Note: The two nuclei just can be separated  in SDSS images, we list magnitudes from SDSS for both components and total magnitudes from other bands for the whole system.\\
      $^{a}$ SDSS model magnitude.\\
     $^{b}$ Magnitudes are for isophotal  fiducial elliptical-aperture.
\end{table}

\section{RESULTS AND DISCUSSIONS}
\subsection{Confirmation of a dual AGN}
Fig.\,1 (c.f. also \S 3.4)  shows clear evidence of interaction between the two components and the chance of the system being a coincidental superposition of two physically unrelated objects or the possibility of being a gravitationally lensed system can both be squarely  ruled out. 
However, images alone are insufficient to distinguish starbursts from AGNs of types 1 or 2.
With the spectra obtained from the {\it HST\/} and LAMOST, we can constrain the nature of the two components of  J0038+4128 by examining the widths of  observed emission lines as well as the locations of the measured line flux ratios on the  line-ratio diagnostic diagrams (Baldwin et al. 1981; Veilleux \& Octerbrock 1987; Kewley et al. 2001, 2006; Kauffmann et al. 2003).
Fig.\,2 shows the {\it HST\/}\, UV  spectrum of J0038+4128S, which clearly exhibits broad emission features identified as H~{\sc i}  Ly$\alpha$ and the Si~{\sc iv} $\lambda\lambda$1394,1403 lines. 
Since Ly$\alpha$ and the N~{\sc v} doublet ($\lambda\lambda$1239,1243) are blended, we fit the observed profile by the sum of two pairs of two Gaussian, one broad and one narrow, with one pair for Ly$\alpha$ and the other for the  N~{\sc v} doublet. When fitting, data points corresponding to most prominent absorption features are masked and excluded from the fitting.
The fitting yields a full width at half maximum (FWHM) for the broad component of Ly$\alpha$ of 4700 $\pm$ 70 km s$^{-1}$ , indicating J0038+4128S harbors a Seyfert 1 type nucleus. 
Furthermore, a close examination of the UV spectra reveals absorption lines that are seen among $\sim$ 50\% Seyfert 1 galaxies  (Crenshaw et al. 1999; Dunn et al. 2007).

J0038+4128N is clearly not a type 1 AGN since the optical spectrum obtained with the LAMOST reveals only narrow emission lines. 
Its nature, whether it is a Seyfert 2, a starburst, or a composite of both can be established by examining the locations of the diagnostic line ratios [O~{\sc iii}] $\lambda$5008/H$\beta$ versus [N~{\sc ii}] $\lambda$6585/H$\alpha$, [S~{\sc ii}] $\lambda\lambda$6718,6733/H$\alpha$ and [O~{\sc i}] $\lambda$6302/H$\alpha$  on the line-ratio diagnostic diagrams.
We measure line fluxes by fitting Gaussians to profiles of detected emission lines, plus a first or second order polynomial for the continuum\footnote{ For the H$\alpha$ and [N~{\sc ii}] doublet, we fit the continuum by a second order polynomial due to the $B$-band absorption to the blue of these three emission lines and possibly the weak contamination from J0038+4128S caused by the high seeing.}.  
From the measured Balmer decrement\footnote{We use an intrinsic, dust-free Balmer decrement of ${\rm H}\alpha/{\rm H}\beta$ ratio of 3.1 for Case B recombination at an electron temperature $T_{\rm e} = 10^4$\,K  and a density  $n_{\rm{e}}$ $\sim$ $10^{2-4} $\,cm$^{-3}$ for Seyfert galaxies (Osterbrock 1989).}, we estimate a colour excess $E(B-V) = 0.71$ for J0038+4128N, a value consistent with the expectation for Seyfert 2 galaxy.
The extinction corrected (for both the host and the foreground Galactic extinction) line ratios of J0038+4128N 
(Fig.\,3) show clearly that it is a Seyfert-type galaxy. 
We use the diagnostic doublet  line ratio of [S~{\sc ii}] $\lambda$6718/$\lambda$6733, assuming a typical electron  temperature of $10^{4}$ K, to estimate the electron density $n_{\rm{e}}$ of the narrow-line region of J0038+4128N. 
The result log  $n_{\rm e}/{\rm cm}^{-3}$ $\sim$ 2.5 is consistent with typical values found for narrow-line regions of Seyfert galaxies.
However, it is still ambiguous whether the gas of the narrow-line region  in J0038+4128N is ionized by  nuclear emission from J0038+4128S or  by that from J0038+4128N itself.
The host galaxy spectrum of J0038+4128N is seen clearly in Fig.\,2, with the Fe~{\sc i} $\lambda$5268 and the Fe~{\sc i} $\lambda$5617 lines well detected.
The redshift derived from the Fe~{\sc i} $\lambda$5617 line of the host galaxy is 0.07351$\pm$0.00011, consistent with the value derived from the narrow emission lines of J0038+4128N (see \S 3.2). 
The signal-to-noise ratio of the spectrum of the host galaxy taken at J0038+4128H with the LAMOST is unfortunately too low to see the underlying absorption features, let alone measurement of the Fe~{\sc i} lines.
Nevertheless, the consistency in redshift of emission lines from the ionized gas and absorption lines from the host galaxy strongly supports that the gas around the nuclear region of J0038+4128N is ionized locally. 
In other words, there are two ionized sources, J0038+4128S and J0038+4128N.

On the other hand, we used Yunnan Faint Object Spectrograph and Camera on Yunnan 2.4 m telescope (pixel size 0\farcs283) to obtain the long-slit medium resolution ($R \sim 2200$, $\lambda = 4970-9830{\rm \AA}$) spectra of J0038+4128 on 2013 November 10.
With clear sky condition and $\sim$ 1\farcs0 seeing, the J0038+418S and J0038+4128N are spatially resolved as shown in Fig.\,4, which confirms that J0038+4128S is a Seyfert 1 galaxy revealed by the {\it HST}\, UV spectrum and J0038+4128N is a Seyfert 2 galaxy revealed by the LAMOST optical spectrum.

In addition, the diagnostic line ratios of J0038+4128H are presented in Fig.\,3 and the results also suggest that J0038+4128H  is consistent with AGN ionization. 
The classical size of a narrow-line region of Seyfert 1\,galaxy is about 1-2 kpc (e.g. Bennert et al. 2006) and corresponding  to 1\farcs3-2\farcs7 in J0038+4128. 
Considering the small ~2\farcs8 distance between the centre of the aperture towards J0038+4128H and J0038+4128S core and the high seeing, the spectra taken at J0038+4128H may mainly come from the narrow-line region of J0038+4128S. 
That is why the measured line ratios at J0038+4128H are consistent with AGN ionization.
But it is a little strange that the J0038+4128H and J0038+4128N share the similar ionization conditions as shown in Fig.\,3.
 Future optical integral-field or long-slit spectroscopy of high spatial resolution are needed to reveal the ionization conditions for the whole system.

In summary, the existing data, both imaging and spectroscopy, strongly suggest that J0038+4128 is a Seyfert 1-Seyfert 2 AGN pair.

\begin{figure}
  \centering
  \includegraphics[width=8.5cm,height=3.1cm]{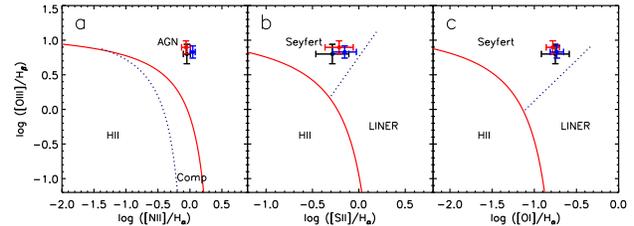}
\caption{ Diagnostic diagrams and measured line ratios from the LAMOST spectra of J0038+4128N (black dot) and J0038+4128H (red dot for 20111024 and blue dot for 20111028). In panel a, the red curve represents the theoretical starburst limit of Kewley et al. (2001) and the blue dotted curve represents the empirical separation between H {\sc ii} regions and AGNs (Kauffmann et al. 2003). H~{\sc ii} regions fall lie below the blue dotted curve, AGNs dominate the region above the red curve, whereas objects of composite H~{\sc ii}-AGN type (Comp) lie between the two curves. In panels b and c, all the curves are taken from Kewley et al. (2006). H~{\sc ii}-regions fall in the area below the red curve, LINERs lie above the red curve but below the blue dotted curve, and Seyferts lie above both the red and blue dotted curves.}
\end{figure}

\begin{figure}
  \centering
  \includegraphics[width=8.4cm,height=4.0cm]{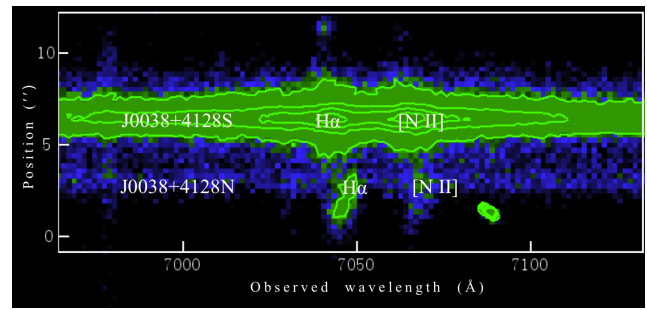}
\caption{Segment of the two-dimensional long-slit spectrum of J0038+4128 that exhibits spatially resolved  broad H${\alpha}$ emission line in the location of J0038+4128S and narrow emission lines (H${\alpha}$+[N~{\sc ii}]) in the location of J0038+4128N. The position of the long-slit is set to cross the centres of J0038+4128N and J0038+4128S.}
\end{figure}

\subsection{Radio and X-ray properties of J0038+4128}

J0038+4128 is detected (but unresolved) at 1.4\,GHz by the NRAO VLA Sky Survey (NVSS; Condon et al. 1998) and at 325 MHz in a radio survey of M31 (Gelfand et al. 2004), with an integrated flux of 5.2 $\pm$ 0.4 mJy and 8.3 $\pm$ 0.72 mJy, respectively. 
The radio spectral index $\alpha$ ($F_{\nu} \propto \nu^{-\alpha}$) is about 0.32, and indicates that J0038+4128 is a compact radio source.
The rest frame radio luminosity at 5 GHz  is estimated at 2.12$\times10^{{39}}$ erg\,s$^{{-1}}$, assuming the above power-law spectral index.
 The rest-frame $B$-band luminosity is about 5.17$\times10^{{43}}$ erg s$^{{-1}}$ estimated from the SDSS $u$ and $g$ magnitudes by the equation of Vanden Berk et al. (2001)\footnote{We modelled the SDSS $u$ and $g$-band images of J0038+4128 using the galaxy structure fitting code GAFIT v3.0 (Peng et al. 2002) by two Gaussian profiles for the dual AGN and a S\'ersic component for the extended host galaxy.}

The radio and optical luminosities thus place J0038+4128 as a typical Seyfert galaxy (Sikora et al. 2007).
J0038+4128 has also been detected in the {\it XMM-Newton} slew survey (Saxton et al. 2008). 
The survey yields  two X-ray sources close to J0038+4128. Of the two sources, the one with the larger position offset has parameter {\it Ver\_Pusp} set to true which means that its position is poorly constrained.
Thus, we have adopted measurements of the X-ray source closest to J0038+4128S (with an offset of only 2\farcs1 from J0038+4128S). The source has an X-ray flux of (1.3 $\pm$ 0.31 $)\times10^{-11}$ erg\,cm$^{-2}$\,s$^{-1}$ at  band 0 (0.2--12 keV) and (4.6 $\pm$ 1.1) $\times10^{-12}$ erg\,cm$^{-2}$\,s$^{-1}$ band 5 (0.2--2 keV), and is not detected at band 4 (2--12 keV). 
Given a Galactic absorption value of $N_{\rm H}$ = 4.4$\times$10$^{20}$\,cm$^{-2}$, estimated from $E(B - V) = 0.069$,  using the relation of Predehl \& Schmitt (1995), we find an X-ray luminosity \footnote{webPIMMS:http://heasarc.nas.gov/Tools/w3pimms.html} $L_{0.5-10\,\rm keV}\sim$1.5 $\times10^{44}$ erg\,s$^{-1}$ for a photon index $\Gamma = 1.8$.
The above X-ray luminosity  and the rest-frame $B$-band absolute magnitude $M_{B} = -20.5$ also show that J0038+4128 is a Seyfert-type galaxy (Brusa et al. 2007).

\subsection{Relative line-of-sight velocity of  the dual AGN}
For J0038+4128N, we measured the redshift by fitting Gaussian profiles to emission lines detected in the LAMOST red arm spectra only (5800--9000\,{\AA}), given the higher accuracy of wavelength calibration of the red arm spectra compared to those of the blue arm (3700--6000\,{\AA}). 
By fitting the [N~{\sc ii}] $\lambda\lambda$6550,6585, H$\alpha$ and the [S~{\sc ii}] $\lambda\lambda$6718,6733, we derive an average redshift of $z_{\rm N} = 0.07328\,\pm\,0.00014$ for J0038+4128N. 
For J0038+4128S, only the broad Ly$\alpha$ can be fitted reliably (the N~{\sc v} and Si~{\sc iv} lines are too weak to obtain reliable redshifts). 
The measured central wavelengths of the broad and narrow components are not the same, which is normal since in most AGNs the Ly$\alpha$ emission line shows an asymmetric profile.
The broad component yields a redshift of $0.07330\,\pm\, 0.00030$. However, from the UV spectrum plotted in Fig.\,2, we find that the narrow component  is closer to the overall  centroid  of the whole Ly$\alpha$ profile, thus adopt the  redshift $z_{\rm S} = 0.07177\,\pm\, 0.00015$  of narrow component of Ly$\alpha$ as the system value of J0038+4128S.
The value is also consistent with the estimate of  Barbieri \& Romano (1976).
From $z_{\rm N}$ and $z_{\rm S}$, we find that the dual AGN nuclei have a relative line-of-sight velocity of 453 $\pm$ 87 km s$^{-1}$.

\subsection{Host galaxy morphology}
With the high spatial resolution images provided by the {\it HST},  the morphology and structure of J0038+4128 can be studied in detail.
In the $F$336$W$/{\it U}-band  image, which reveals the star-forming activities, two pairs of bi-symmetric spiral arms are detected in a binary AGN system for the first time.
The results are consistent with the numerical simulations of Di Matteo et al. (2005).
They have simulated the merging of two discs galaxies of the size of the Milky Way and found the tidal interactions can distort the discs into a pair of bi-symmetric spirals as the two galaxies begin to coalesce. 
In addition to the bi-symmetric spiral arms, there are several compact knots scattered around the two nuclei.
Knots triggered by interaction are common features in merging galaxies (Villar-Mart\'in et al. 2011).
The ongoing process of star formation  in those knots can be confirmed by further spectroscopy.
In the $F$555$W$/{\it V}-band image, we see a stream-like substructure along the north-western edge. 
This may indicate that the two galaxies may have encountered more than once.

\subsection{Variability in the optical and infrared bands}
J0038+4128 is found to be a fast variable in the optical , as shown by Barbieri \& Romano (1976).
They detect irregular variations of large amplitudes ($\sim$ 1.5 mag) on short time-scales (few days) in $B$-band based on hundreds of  measurements accumulated with the 67/92 cm Schmidt Telescope of Asiago and the 182 cm Telescope of Cima Ekar from 1965 September 5 to 1975 January 13.
The strong, fast optical variability indicates the presence of a fast flare component (tens of days) in the light curve of  J0038+4128. 

Finally, J0038+4128 is also detected by the {\it Wide-Field Infrared Survey Explorer} ({\it WISE}; Wright et al. 2010) in all bands. 
{\it WISE} maps the entire sky at 3.4, 4.6, 12, and 22 $\mu$m (bands  $W1$, $W2$, $W3$ and $W4$, respectively) at an angular resolution of 6\farcs1, 6\farcs4, 6\farcs5\ and 12\farcs0, respectively.
In the {\it WISE} All-Sky Data Release Source Catalog, a variability flag (of integer values  0--9), {\it var\_flg}, is assigned to every detected object in each band, indicating the probability of possible flux variations. 
The greater the value of {\it var\_flg}, the higher the possibility of variability (see Hoffman et al. 2012 for detail).
For J0038+4128, the flag has the highest value of 9 in both $W1$ and $W2$ bands and the light curves in the two bands show variability of amplitude of about 0.15 mag over a time-scale of half year.

\section{SUMMARY}
J0038+4128 is resolved to contain two compact nuclei in both $F$336$W$/{\it U}- and $F$555$W$/{\it V}-band images secured with the {\it HST} WFPC2.
The {\it HST} GHRS UV spectrum of the southern nucleus shows broad Ly$\alpha$ emission (FWHM $\sim$ 4700\,kms$^{-1}$), which indicates that it is a Seyfert 1 galaxy.
The LAMOST optical spectrum of the northern nucleus shows narrow emission lines only. Line diagnostics as well as the  consistency in redshift between emission lines from the ionized gas and absorption lines from the host galaxy suggest it is a Seyfert 2.
Therefore, the {\it HST} and LAMOST data confirm that J0038+4128 is a  Seyfert 1-Seyfert 2 AGN pair with a projected spatial separation of  4.7 kpc (3\farcs44) and a line-of-sight relative velocity of 453 km s$^{-1}$.

 The $F$336$W$/{\it U}-band image of J0038+4128, also reveals for  the first time for a dual AGN system two pairs of bi-symmetric spiral arms. 
 The {\it HST}\ images also reveal a number of compact star-forming knots as well as some tidal stream features.
 
 Future optical integral-field spectroscopy of high spatial resolution would be extremely useful for further study of  J0038+4128, in order to reveal the physical conditions and chemical properties of both the nuclear ionized gas and gas of the host galaxy for the whole entire system.
 Such a study  will provide us much needed information of the co-evolution of the host galaxy and the central black holes.
 
 \section*{Acknowledgements} 
 This work is supported by National Key Basic Research Program of China 2014CB845700.  We thank Professor Fukun Liu and Dr. Andreas Schulze for providing valuable comments and suggestions of this paper. 
 
This work has made use of data products from the Large Sky Area Multi-Object Fibre Spectroscopic Telescope (LAMOST), Sloan Digital Sky Survey (SDSS), {\it Galaxy Evolution Explorer} ({\it GALEX}), Two Micron All Sky Survey (2MASS), {\it Wide-field Infrared Survey Explorer} ({\it WISE}), NRAO VLA Sky Survey (NVSS),  {\it XMM-Newton\/} and Yunnan 2.4 m telescope.

\end{document}